\documentclass[journal]{IEEEtran}

\usepackage{ifpdf}
\usepackage{cite}
\usepackage{subfigure}
\usepackage{graphicx}
\DeclareGraphicsExtensions{.pdf,.jpg,.png}

\usepackage[cmex10]{amsmath}
\usepackage{amssymb}
\usepackage{array}
\usepackage{mdwmath}
\usepackage{mdwtab}
\usepackage{pslatex}
\usepackage{color}
\usepackage{textcomp}
\usepackage{placeins}
\usepackage{eqparbox}
\usepackage{url}
\usepackage{stfloats}
\usepackage{multicol}
\usepackage{subfigure}
\usepackage{listings}
\usepackage{balance}
\usepackage{footnote}

\usepackage{float}
\usepackage{algorithmicx}
\usepackage{algpseudocode}
\usepackage{gensymb}

\begin{document}

\title{\vspace{-3mm}\textcolor{black}{\huge On Unified Vehicular Communications and Radar Sensing in Millimeter-Wave and Low Terahertz Bands}}

\author{\vspace{-1mm}
\IEEEauthorblockN{Vitaly Petrov\IEEEauthorrefmark{1}, Gabor Fodor, Joonas Kokkoniemi, Dmitri Moltchanov, Janne Lehtom{\"a}ki,\\Sergey Andreev,
Yevgeni Koucheryavy, Markku Juntti, Mikko Valkama\vspace{-5mm}}
\thanks{\IEEEauthorrefmark{1}A part of this work has been completed during the research visits of Vitaly~Petrov to Ericsson Research, Sweden.}
\thanks{G. Fodor is with Ericsson Research and the School of Electrical Engineering and Computer Science of KTH Royal Institute of Technology, Sweden.}
\thanks{V. Petrov, D. Moltchanov, S. Andreev, Y. Koucheryavy, and M. Valkama are with Tampere University, Finland.}
\thanks{J. Kokkoniemi, J. Lehtom{\"a}ki, and M. Juntti are with the Centre for Wireless Communications, University of Oulu, Oulu, Finland.}
\thanks{This work was supported in part by Academy of Finland 6Genesis Flagship (grant 318927) and by the framework project TAKE-5: The 5th Evolution Take of Wireless Communication Networks, funded by Tekes. V.~Petrov acknowledges the support of HPY Research Foundation funded by Elisa.}
\thanks{\copyright\,\,2019 IEEE. The work has been accepted for publication in IEEE Wireless Communications, 2019. Personal use of this material is permitted. Permission from IEEE must be obtained for all other uses, including reprinting/republishing this material for advertising or promotional purposes, collecting new collected works for resale or redistribution to servers or lists, or reuse of any copyrighted component of this work in other works.}
}

\maketitle

\begin{abstract}
Future smart vehicles will incorporate high-data-rate communications and high-resolution radar sensing capabilities operating in the millimeter-wave and higher frequencies. These two systems are preparing to share and reuse a lot of common functionalities, such as steerable millimeter-wave antenna arrays. Motivated by this growing overlap, and advanced further by the space and cost constraints, the vehicular community is pursuing a vision of unified vehicular communications and radar sensing, which represents a major paradigm shift for next-generation connected and self-driving cars. This article outlines a path to materialize this decisive transformation. We begin by reviewing the latest developments in hybrid vehicular communications and radar systems, and then propose a concept of unified channel access over millimeter-wave and higher frequencies. Our supporting system-level performance characterization relies upon real-life measurements and massive ray-based modeling to confirm the significant improvements brought by our proposal to mitigating the interference and deafness effects. Since our results aim to open the door to unified vehicular communications and radar sensing, we conclude by outlining the potential research directions in this rapidly developing field.
\end{abstract}

\vspace{-3mm}
\section{Today's Vehicular Communications and Radars}\label{sec:Introduction}

Over the last decade, vehicular communications have experienced a fundamental transformation by absorbing the impact of several waves of research and development trends in this field. \emph{The first wave} came in the mid-2000s, with the IEEE 802.11p initiative that enabled data exchange between vehicles over the unlicensed frequency band around $5.9$\,GHz. The subsequent IEEE 802.11p-2010 standard amendment supported ad-hoc vehicular communications with data rates of up to $27$\,Mbit/s over $10$\,MHz-wide channels and was later superseded by conventional Wi-Fi IEEE 802.11-2012 specification. In parallel, vehicle-centric operations have been facilitated by additional standards that regulated the higher layers of a vehicular network (e.g., IEEE 1609.32 Standard for Wireless Access in Vehicular Environments, WAVE)~\cite{dsrc_survey}. This formed a consistent set of specifications for microwave-based vehicular communications, referred to as cooperative, connected, and automated mobility (CCAM) in Europe and dedicated short-range communications (DSRC) in the United States.

\begin{figure}[t!]
\centering
\includegraphics[width=1.0\columnwidth]{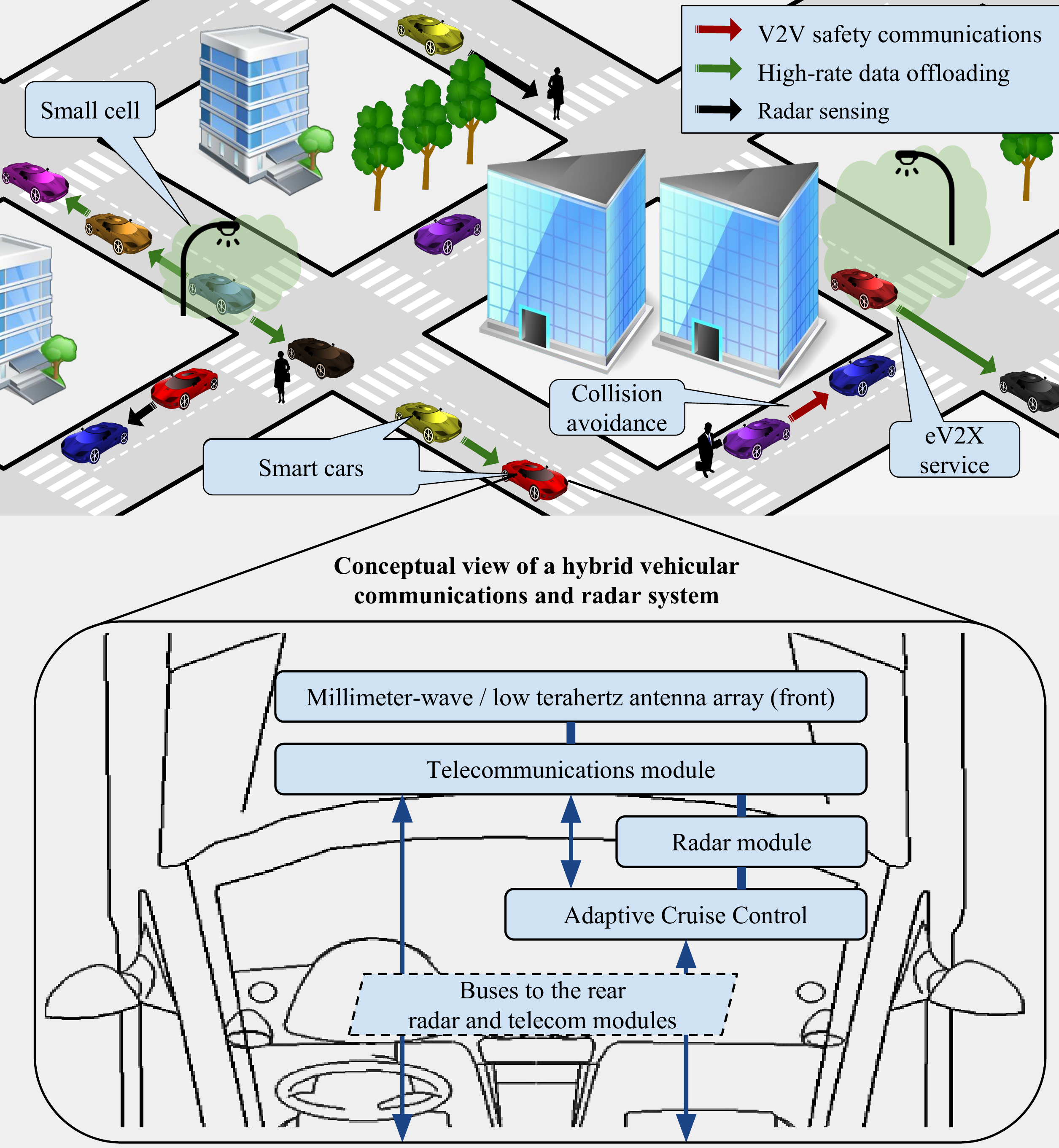}
\vspace{-5mm}
\caption{\textcolor{black}{Joint V2V communications and automotive radar applications.}}
\label{fig:v2v_comm}
\vspace{-5mm}
\end{figure}

By design, the 802.11p-family of technologies is dedicated to direct vehicle-to-vehicle (V2V) interactions. Hence, these are not intended to be used for vehicle-to-pedestrian (V2P) or vehicle-to-infrastructure (V2I) data exchange. In response to this gap, \emph{the second wave} of standardization activities has been dedicated to vehicular-to-everything (V2X) communications over 3GPP LTE technology, where uplink, downlink (for V2P and V2I), and sidelink (for V2V) operations were ratified~\cite{gabor1}. Early work was initiated in 2014, while the key requirements and use cases were identified by the \textcolor{black}{3GPP's Technical Specification Group on Service and System Aspects (TSG-SA)} in 2016. The corresponding enablers to support mission-critical V2X communications were presented in 2017 as part of LTE Release 14. V2X over LTE enables reliable connectivity within hundreds of \textcolor{black}{meters at around $100$\,ms end-to-end latencies (from the source to the destination application)}.

\vspace{-2mm}
\subsection{Emerging mmWave V2X Communications and Radar}

Within LTE Release 15 and beyond, 3GPP explores the feasibility of next-generation (5G-grade), enhanced V2X (eV2X) use cases, which range from platooning to remote driving~\cite{my_white_paper_automotive_vision}. Many of these scenarios require extremely high data rates and reliability levels, as well as low latencies~\cite{petrov_jsac_sdn}. For example, ``bird's eye view'' and ``see-through vision'' demand up to $50$\,Mbit/s at $50$\,ms delay, while ``automated overtake'' requires $10$\,ms latency with $10^{-5}$ reliability. These stringent requirements engage the community to initiate \emph{the third wave} of V2X communications research, by targeting the New Radio (NR) technology that will complement LTE~\cite{gabor2}. It is expected to primarily operate in millimeter-wave (mmWave) spectrum; that is, $30$--$300$\,GHz. NR-V2X will augment LTE V2X in massive deployments of connected vehicles and its \textcolor{black}{standardization has started recently}. The academic groups are advancing this vision by exploring mmWave-based V2X communications over $60$\,GHz, $79$\,GHz, and \textcolor{black}{even low terahertz frequencies at $300$\,GHz and beyond~\cite{heath_v2v_sensing}.}

Today's automotive radar sensing systems also occupy the mmWave band and, particularly, the frequencies of $76$--$81$\,GHz. The existing solutions can generally be classified into three categories: (i) long-range radars (up to $100$\,m) that typically sense the forward direction for the adaptive cruise control applications; (ii) middle-range radars (tens of meters) dedicated to cross-traffic alerts or blind-spot detection; and (iii) short-range radars (several meters), which are exploited by parking assistance and pre-crash services. Solutions operating in the $77$ and $79$\,GHz bands are already in commercial use, while the total number of mmWave radars can easily reach $10$ for a modern high-class vehicle.

\vspace{-2mm}
\subsection{From Trial Deployments to Massive mmWave V2X}

While the \emph{feasibility} of mmWave V2X communications and radar sensing has already been confirmed, the focus of recent research efforts has shifted towards enabling \emph{scalable} operation in massive (semi-)autonomous driving with smart interconnected vehicles (see Fig.~\ref{fig:v2v_comm})\textcolor{black}{~\cite{ref_link1}}. The major scalability concerns are -- jointly for data transmission and radar sensing -- related to massive interference from neighboring vehicles, which challenges the reliability, latency, and achievable data rate of NR-V2V communications~\cite{heath_v2v_sensing}. Further, there may be dangerous sensing inaccuracies that lead to the emergence of ``ghost'' obstacles. The level of interference in such systems \emph{must} be kept under control \textcolor{black}{to make collective autonomous driving safe and efficient}.

In this article, we tackle the aforementioned challenges by advancing the paradigm of \emph{hybrid vehicular communications and radar sensing} over the mmWave band and beyond. We systematically review and classify the contemporary approaches in this field as well as propose a conceptual joint channel access framework that aims to reduce the levels of interference in both communications and radar operations. We evaluate this concept in a realistic urban scenario by relying on: (i) real-life measurements and (ii) our in-house ray-based modeling framework to quantify the potential gains of unifying mmWave-based vehicular communications and radar sensing. Our proposal can be integrated into future releases of 3GPP NR and IEEE 802.11. We conclude by outlining the research perspectives in this area that are meant to open a discussion on the future mmWave and beyond vehicular networking.

\section{\hspace{-1.55mm}Review on Joint Telecom \& Radar for V2V}
\label{sec:review}

The concept of joint communications and radar applications has been evolving continuously over several decades. More recently, utilization of mmWave frequencies in this context became particularly beneficial, since much of the communications and radar systems functionality partially overlaps in these bands. In this section, we review the contemporary approaches to hybrid vehicular communications and radar sensing, which typically follow either of the \emph{three} options:

\vspace{-2mm}
\subsection{Time-Domain Duplex (TDD)}
\label{sec:tdd_class}

The first approach assumes independent operation of telecom and radar subsystems, while the utilization of antenna hardware may be shared between them in a time-based manner~\cite{tdd_system}. There is a switching process between the two time intervals: (i) when the antenna system (e.g., a single antenna or an antenna array) is used for data communications and (ii) when the antenna system is exploited to transmit and receive radar sequences. The corresponding modulated waveform can be generated with a direct digital synthesizer (DDS)~\cite{tdd_system}.

TDD class of solutions features lower implementation complexity. At the same time, the resultant system scalability may be rather limited, since fast switching between data and radar modes cannot be synchronized effectively among all of the communicating vehicles. Hence, when one of the cars is transmitting its data, the destination node may reside in the radar mode, not being able to receive it. In addition, since perfect time synchronization is difficult to achieve, the data and radar transmissions can interfere with each other.

\vspace{-2mm}
\subsection{Telecom Messages over Radar Transmissions (ToR)}

The second approach suggests modulating the data messages on top of the radar transmissions. This category typically employs Pulse Position Modulation (PPM) for the data transmissions. The corresponding systems are featured by relatively high radar resolution and accuracy since no stringent constraints on the choice of the radar sequences are imposed, while each transmission serves the purpose of sensing.

The scalability of these solutions is higher as compared to the TDD schemes since data transmissions do not impede the sensing process of the radar module. However, spectrum utilization remains inherently low, \textcolor{black}{since the choice of applicable modulation and coding schemes is limited to a set of basic solutions}. As a result, the achievable data rate is low even at the mmWave band: about $200$\,Mbps over $3$\,GHz bandwidth~\cite{tor_system}.

\begin{figure*}[t!]
\centering
\vspace{-11mm}
\includegraphics[width=1.0\textwidth]{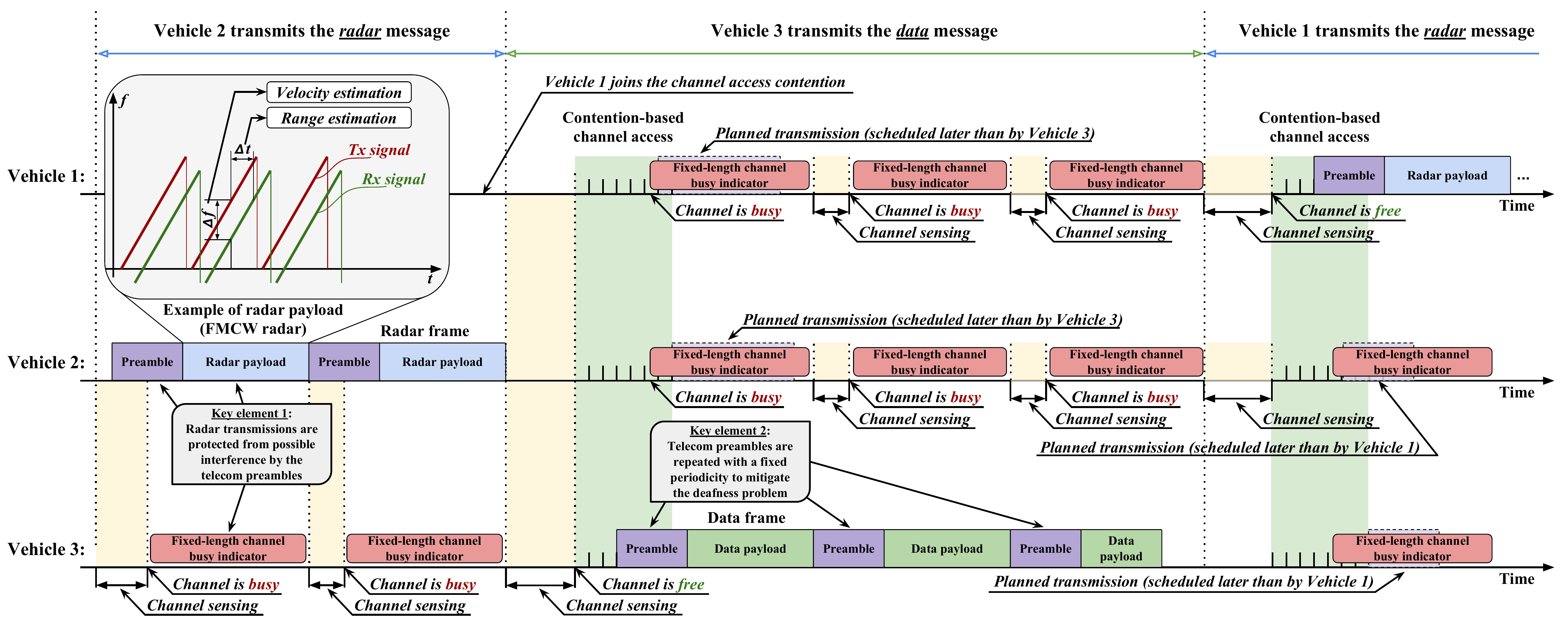}
\vspace{-7mm}
\caption{\textcolor{black}{Proposed integration of V2V communications and radar.}}
\label{fig:concept}
\vspace{-4mm}
\end{figure*}

\vspace{-2mm}
\subsection{Radar Sensing over Telecom Transmissions (RoT)}

The solutions in this group are more recent and specific to mmWave frequencies. Their feasibility is maintained by the specialized properties of the preambles used in mmWave radio communications standards (e.g., IEEE 802.11ad/ay or 3GPP NR). This approach suggests reusing the same preamble detection mechanisms in order to provide distance and velocity estimation for the nearby driving vehicles.

It has recently been validated experimentally and shown to be beneficial from both the data rate and the sensing accuracy perspectives~\cite{rot_system_heath_tvt}. At the same time, the sensing capabilities of the developed prototype are still inferior to those offered by the commercially-available dedicated radar systems as well as remain vulnerable to uncontrolled interference from other cars initiating their data transmissions nearby. Further, RoT approaches require tight agreement between many stakeholders as they embrace the know-how of radars from vehicular vendors -- all the relevant signaling has to be standardized publicly. Hence, while this class of systems has technical advantages as compared to the previous two, the business perspectives of the corresponding solutions remain unclear.

\section{Harmonized V2V Telecom \& Radar Framework}
\label{sec:concept}

Summarizing our above review of existing approaches for joint radar and comms in V2V, current functionality of telecom and radar subsystems is mostly separated, as the associated modules operate \emph{independently} and only occasionally share common equipment in the transceiver to access the radio channel. Addressing the gap, we seek to explore the synergy of the involved subsystems rather than aim at their straightforward blending within a single physical device.

\vspace{-2mm}
\subsection{Turning Competitors into Allies}

Instead of constructing a communications system on top of the radar functionality or achieving the radar goals over a communications module, we envisage a possibility to \emph{intelligently combine the communications and radar subsystems} in a way that \textcolor{black}{they perform harmonized operations}, while both benefiting from the capabilities of the mmWave band. \textcolor{black}{Particularly, we conceptualize a TDD-inspired solution (see subsection~\ref{sec:tdd_class}) where synchronization between vehicular communications and radar operations is conducted not within a single car but rather propagates across the entire collision domain of the neighboring vehicles.}\footnote{\textcolor{black}{Here, ``collision domain'' refers to the set of vehicles potentially affecting each others' transmissions: the generated interference is higher than the noise level.}} The communication and radar functionalities \textcolor{black}{need} to be either placed into a unified module or to be located as separate but interconnected devices, still employing a certain level of hardware reuse, as well as synchronization to share the common wireless channel in the time domain~\cite{tdd_system}. \textcolor{black}{The solution reduces the probability that the radar sensing performed by one of the vehicles interferes with the transmissions coming from another vehicle and vice versa.}

Our proposal offers the much needed \textcolor{black}{scalability and reliability as well as} flexibility for vendors to develop their own radar waveforms without losing the integrity within the employed communications protocol. Moreover, with this approach, the accuracy of the radar subsystem can also be improved by integrating the results of the communications preamble detection and the actual radar waveform operation~\cite{rot_system_heath_tvt}. We thus advocate for a harmonized framework over a multi-tenancy multiple access channel, \textcolor{black}{where the same preambles are used to indicate radar as well as data transmissions. Hence,} the radar messages avoid interference by following well-established random-access communications mechanisms, whereas the notorious deafness problem is resolved by \textcolor{black}{periodic transmissions of additional preambles} (see Fig.~\ref{fig:concept}). The resulting \textcolor{black}{Radar-Aware} Carrier-Sense Multiple Access (RA-CSMA) scheme comprises two major components: (i) a unified frame structure for transmitted communications and radar messaging and (ii) a radar-aware \textcolor{black}{preamble detection based} channel-access protocol for directional V2V communications.

\vspace{-2mm}
\subsection{Unified Frame Structure}

In the proposed system design, the communications and radar module controls the shared antenna system. \textcolor{black}{The vehicle performing the radar sensing first utilizes a telecom module to participate in contention-based channel access and, as a result of this procedure, transmits a preamble to ``reserve'' the channel for the duration of a frame (see Fig.~\ref{fig:concept}). Then, the vehicle changes its operating mode from communications to radar to initiate the actual sensing. The sensing period may consist of a single or multiple transmissions/receptions of the radar sequences, as illustrated in Fig.~\ref{fig:concept} (top left). Finally, once the ``reserved'' time is over, the vehicle switches its operating mode back to communications and either ``reserves'' the channel again by transmitting another preamble or switches to the Rx mode and continues transmitting/receiving other data by following the CSMA-based access procedure.}

In order to implement the intended operation, an additional fixed-length field to indicate the frame category has to be introduced. It needs to be located in the header to enable fast decoding of the frame category at the hardware level. All of the frames coming from other vehicles and marked as ``radar'' are immediately dropped at the receiver. The described operation allows for utilizing a unified collision avoidance mechanism for both the communications and radar subsystems, hence reducing the time-to-market and manufacturing costs. Moreover, it offers robust protection of the radar sequences, since radar frame transmissions enjoy ``collision-free'' environment: \textcolor{black}{following the described procedure, no other vehicles in the same collision domain can initiate a transmission as long as the medium is reserved}. The latter is crucial for developing safety solutions as it \textcolor{black}{reduces interference and thus enables more} accurate distance and obstacle velocity estimation\textcolor{black}{~\cite{Gameiro2018}}.

\vspace{-1mm}
\subsection{Radar-\textcolor{black}{Aware} Channel Access \textcolor{black}{with Preamble Repetition}}

The deafness problem in V2V communications may affect the transmission of both the data and radar messages. The conventional mechanisms based on energy detection or \textcolor{black}{frame header} decoding are not sufficiently robust to rely upon when detecting transmissions of the neighboring vehicles, especially across the reflected paths. To address this challenge, we propose the use of \textcolor{black}{additional} preambles by injecting them into the channel with a fixed periodicity and minimal interframe spacing as illustrated in Fig.~\ref{fig:concept}. It has been demonstrated that preamble-based detection is much more reliable even in extremely low-SNR regimes~\cite{rot_system_heath_tvt}. 

Design of a particular preamble sequence to accommodate the described concept remains an open research issue. It is, however, worthwhile to note that the discussed preamble detection over reflected/scattered signals is technically feasible since this procedure is already a part of the basic \textcolor{black}{communications and} radar functionality. In \textcolor{black}{the described} channel access scheme, the vehicle that is about to transmit first listens on the channel for one inter-preamble interval. If a preamble is detected, the channel busy indicator is set for another interval. Otherwise, the channel is considered idle and the vehicle initiates its channel access procedure. With the proposed approach, the rest of the radio technology signaling can be applied to the envisioned operation with minimal changes.

\textcolor{black}{The outlined channel access procedure introduces an additional delay in the radar sensing operation. However, accounting for $5$\,$\mu$s-long slots (as in IEEE 802.11ay) and the maximum backoff counter of $1024$, the total delay value does not exceed $5.12$\,ms. Furthermore, since the size of the collision domain in realistic deployments is around $2$--$3$ vehicles, the said counter does not typically exceed $16$. This keeps the channel access delay under a few milliseconds, which is in-line with the requirements for beyond-5G mission-critical communications~\cite{petrov_jsac_sdn}. Importantly, a vehicle can travel not more than a few centimeters during such ms-scale intervals. Hence, the extra delay caused by channel access does not affect the radar operations significantly.}

In the following section, we comprehensively study the implications of the proposed RA-CSMA scheme on the performance of mmWave-based V2V applications.

\section{Performance Expectations in mmWave V2V}

\begin{figure}[t!]
\centering
\vspace{-2mm}
\includegraphics[width=1.0\columnwidth]{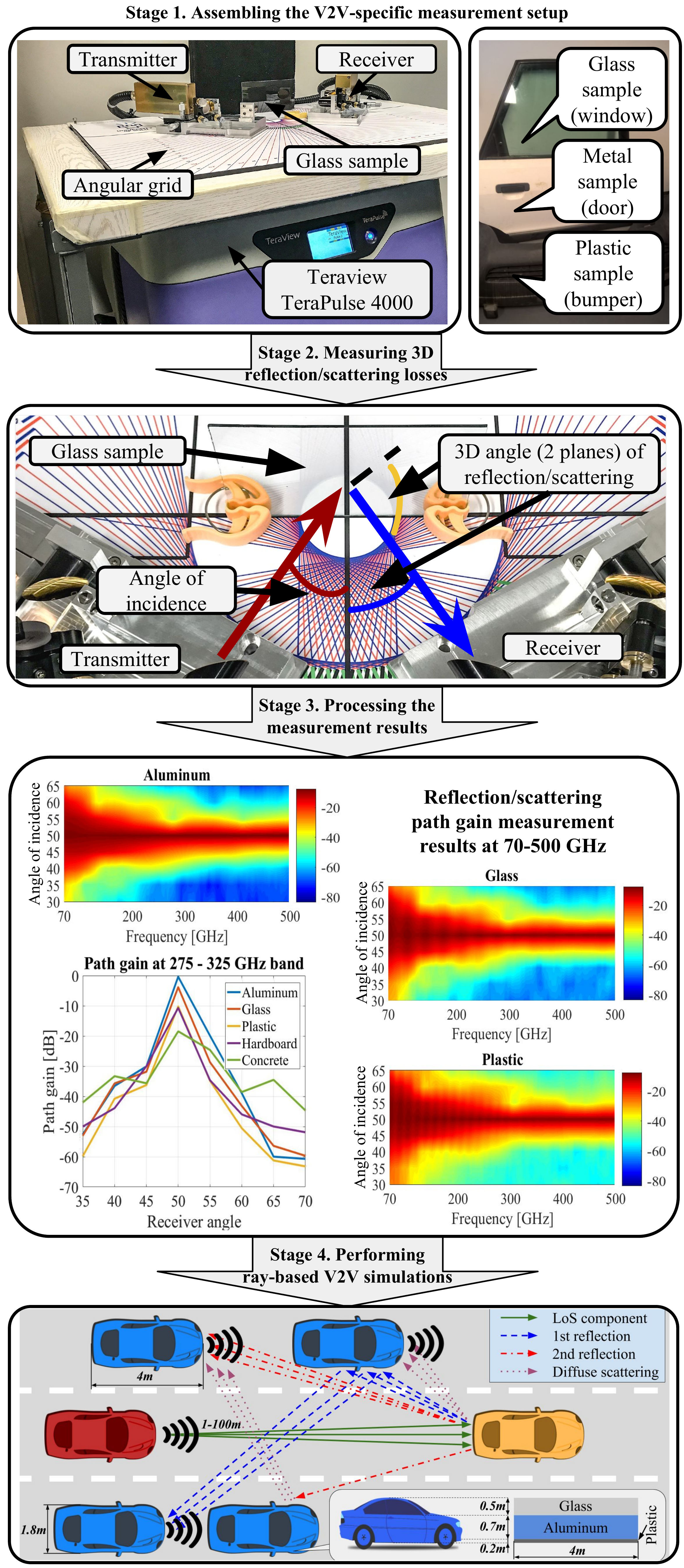}
\vspace{-5mm}
\caption{\textcolor{black}{Our compound measurement--simulation evaluation methodology for the performance assessment of hybrid V2V systems.}}
\vspace{-7mm}
\label{fig:measurement}
\end{figure}

Our proposed RA-CSMA scheme relies upon the ability of the vehicle that initiates a transmission to sense active transmissions by other vehicles in the same collision domain even in cases where only reflected or scattered paths are available at its transceiver. In order to evaluate the proposed concept, we first have to carefully characterize the specifics of mmWave propagation in vehicular environments, with the main focus on the effects related to the mmWave signal reflection and scattering from the bodies of vehicles. For this purpose, an integrated measurement--simulation campaign has been conducted (see Fig.~\ref{fig:measurement}).

\vspace{-2mm}
\subsection{Specifics of mmWave Propagation in V2V Scenarios}
\label{subsec:propagation}

To characterize mmWave signal propagation in realistic V2V environments, we first conduct a measurement campaign that aims to analyze and report on the properties of reflection and scattering at the frequencies of up to $500$\,GHz from the materials typical for the vehicle bodies, including aluminum, glass, and hard plastic \textcolor{black}{(see Fig.~\ref{fig:measurement}, Stage 1)}. The measurements were performed at the University of Oulu by using TeraView TeraPulse $4000$, which is capable of mmWave and terahertz band transmission/reception from $60$\,GHz to $4$\,THz. The signal attenuation has been measured as a function of the selected material and three 2D angles: the angle of incidence and the angles of reflection/scattering in two planes.

To this end, Fig.~\ref{fig:measurement} displays an example of the measurement results for $50\degree$ angle of arrival. Aluminum is the best reflector out of the considered materials with the reflection loss being on the order of $0$--$5$\,dB. In contrast, other materials (glass and, especially, plastic) are characterized by much higher scattering losses, which are particularly visible at the frequencies of above $100$\,GHz. Similar measurements have been carried out for other angles of arrival with the step of $5\degree$ and stored in a way that the ray-based framework can calculate the corresponding path gain as a function of the angle of incidence and 3D angle of reflection/scattering.

\begin{table}[!b]
\vspace{-2mm}
\caption{Parameters of our ray-based study.}
\label{tab:params}
\begin{center}
\vspace{-3mm}
\begin{tabular}{|p{0.15\columnwidth}|p{0.75\columnwidth}|}
\hline
Deployment & \textbf{Area of interest:} Segment of a street, $200$\,m long, $3$ lanes in each direction ($6$ lanes in total)\\
& \textbf{Lane width:} $2.75$\,m\\
& \textbf{Sidewalk width:} $3$\,m\\
& \textbf{Total street width:} $22.5$\,m\\
& \textbf{Propagation:} Urban canyon, street is surrounded by concrete buildings of $30$\,m height\\
\hline

Radio part & \textbf{Environment:} Air, T = $296$\,K and $1.8$\% of water vapor\\
& \textbf{Frequency:} $300$\,GHz\\
& \textbf{Bandwidth:} $10$\,GHz\\
& \textbf{Tx power:} $0$\,dBm\\
& \textbf{Tx antenna gain:} $30$\,dB\\
& \textbf{Rx antenna gain:} $30$\,dB\\
& \textbf{Antennas:} In front and rear bumpers ($0.2$\,m altitude). \textcolor{black}{Perfect beam alignment between Tx and Rx is assumed.}\\
\hline
Vehicles & \textbf{Model:} Parallelepiped $4$\,m $\times$ $1.8$\,m $\times$ $1.4$\,m\\
& \textbf{Material:} Glass (top $0.5$\,m); steel (middle $0.7$\,m); plastic (bottom $0.2$\,m)\\
& \textbf{Reflection/scattering properties:} Defined by the measurement results in subsection~\ref{subsec:propagation}\\
& \textbf{Inter-vehicle distance:} \textcolor{black}{\underline{Setup 1}: Random (exponentially distributed); \underline{Setup 2}: Constant for all the vehicles}\\
& \textbf{Speed:} \textcolor{black}{\underline{Setup 1}: Normally distributed with the mean of $30$\,km/h; \underline{Setup 2}: Constant for all the vehicles, $5$\,km/h}\\
\hline 
Channel access & \textbf{Idealistic:} Perfect TDMA, ideal time synchronization, thus no inter-vehicle interference (upper bound)\\
& \textbf{RA-CSMA:} Proposed scheme, CSMA with preamble repetition (see subsection~\ref{sec:concept} for details) \\
& \textbf{Uncoordinated:} Uncoordinated random access\\
& \textbf{Adaptive access:} Uncoordinated random access with binary exponential backoff operation (from $8$ to $1024$)\\
\hline
\end{tabular}
\end{center}
\end{table}

The obtained measurement results are then applied to evaluate the performance of mmWave V2V communications with different channel access schemes (including the proposed RA-CSMA) in a typical urban deployment. Particularly, a wide city avenue with a randomized deployment of vehicles has been considered (see Fig.~\ref{fig:measurement}, Stage 4). The utilized parameters are summarized in Table~\ref{tab:params}. 

To characterize the introduced scenario and collect first-order performance results, an in-house ray-based modeling framework was employed~\cite{petrov_commag_thz}. It is mainly implemented in Python, \textcolor{black}{operates in a time-driven manner}, and adopts the measurement data from TeraView test environment as input for the reflection/scattering properties of incorporated materials. Our framework utilizes a ray-tracing approach with surface tessellation\textcolor{black}{, where the signal propagation in the air is modeled as proposed in~\cite{jornet_channel}.} For the sake of better accuracy in the output results, all of the acquired intermediate data have been averaged over $1,000$ replications.

\vspace{-2mm}
\subsection{Scalability of Channel Access Schemes for mmWave V2V}
\label{sec:results}

\begin{figure}[h!]
\centering
\includegraphics[width=1.0\columnwidth]{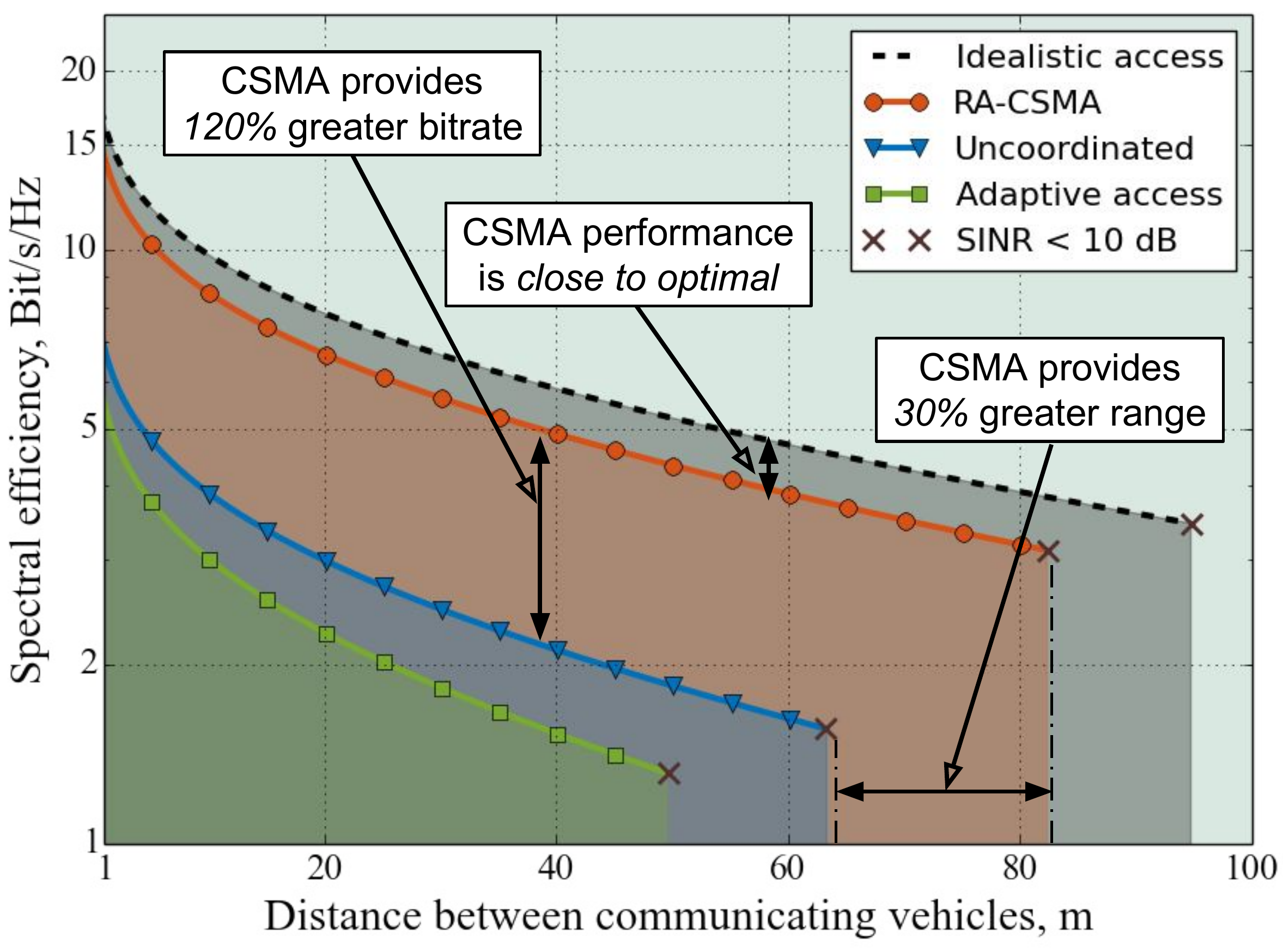}
\vspace{-3mm}
\caption{Spectral efficiency of mmWave V2V system with RA-CSMA.}
\label{fig:capacity}
\end{figure}

First, Fig.~\ref{fig:capacity} reports on the spectral efficiency of a V2V communications link in case the distance between two connected vehicles varies from $1$ to $100$\,m, while all other vehicles are deployed randomly with the average distance of $10$\,m between each other. \textcolor{black}{The inter-vehicle distance is distributed exponentially, as in \underline{Setup 1}} (see details in Table~\ref{tab:params}). The utilization of the envisioned RA-CSMA channel access scheme improves both the link spectral efficiency and the achievable communications range where data transmissions remain sufficiently reliable ($SINR \geq 10$\,dB). It is also observed that with our proposed solution the realistic communications range can reach up to a hundred of meters even with $10$\,GHz of bandwidth and over $300$\,GHz carrier frequency, thus enabling a number of attractive rate-hungry applications currently envisioned for e.g., NR-V2V. Similar RA-CSMA performance is observed for $79$\,GHz and $150$\,GHz mmWave frequencies.

Further, Fig.~\ref{fig:sinr} illustrates the impact of correct preamble detection on the reliability of mmWave-band V2V communications \textcolor{black}{and radar sensing}. This figure demonstrates the average SINR at the receiver as a function of the mean distance between the vehicles in our scenario of interest. Here, the distance between the \textcolor{black}{source} and the \textcolor{black}{destination} vehicles (red and yellow, respectively, in Fig.~\ref{fig:measurement} Stage 4) \textcolor{black}{is modeled similarly to the distances between other neighboring vehicles. It follows one of the two setups that correspond to different deployment scenarios. The first one characterizes a freeway scenario, where the distances between the vehicles are modeled as independent and exponentially distributed random variables. In the second setup, the distances between all of the vehicles are kept constant, thus capturing dense urban traffic.}

\vspace{-3mm}
\subsection{Interference- vs. Noise-Limited Operation of mmWave V2V}

Analyzing Fig.~\ref{fig:sinr}, we emphasize three distinct regimes for mmWave V2V communications \textcolor{black}{and radar sensing}, which are generally agnostic to the selected channel access scheme \textcolor{black}{as well as the assumed distribution of inter-vehicle distances}:

\subsubsection{The ultra-dense regime} corresponds to $<1$\,m of inter-vehicle separation and is characterized by the highest level of SINR. Radio signal attenuation in this case is low, while the impact of interference is marginal due to substantial blockage by car bumpers and wings.

\subsubsection{The dense regime} is characterized by the inter-vehicle distance from $\approx1$\,m to $10$--$20$\,m. The role of interference becomes of primary importance, since the useful signal is not as strong, whereas there are many other vehicles nearby.

\subsubsection{The sparse regime} has the average inter-vehicle distance of greater than $10$--$20$\,m. It is mostly affected by the noise level; the useful signal becomes notably attenuated, while the population of interfering vehicles remains low.

\textcolor{black}{A minor improvement in the SINR when switching from \underline{Setup 1} to \underline{Setup 2} is explained by the fact that the first setup allows for situations where the distance between the communicating vehicles is long, while the distances between the vehicles on the neighboring lanes are short, so there are multiple vehicles to interfere with an active transmission.}

\vspace{-5mm}
\textcolor{black}{\subsection{Reliability Aspects of Radar Sensing}}

\begin{figure}[h!]
\centering
\includegraphics[width=1.0\columnwidth]{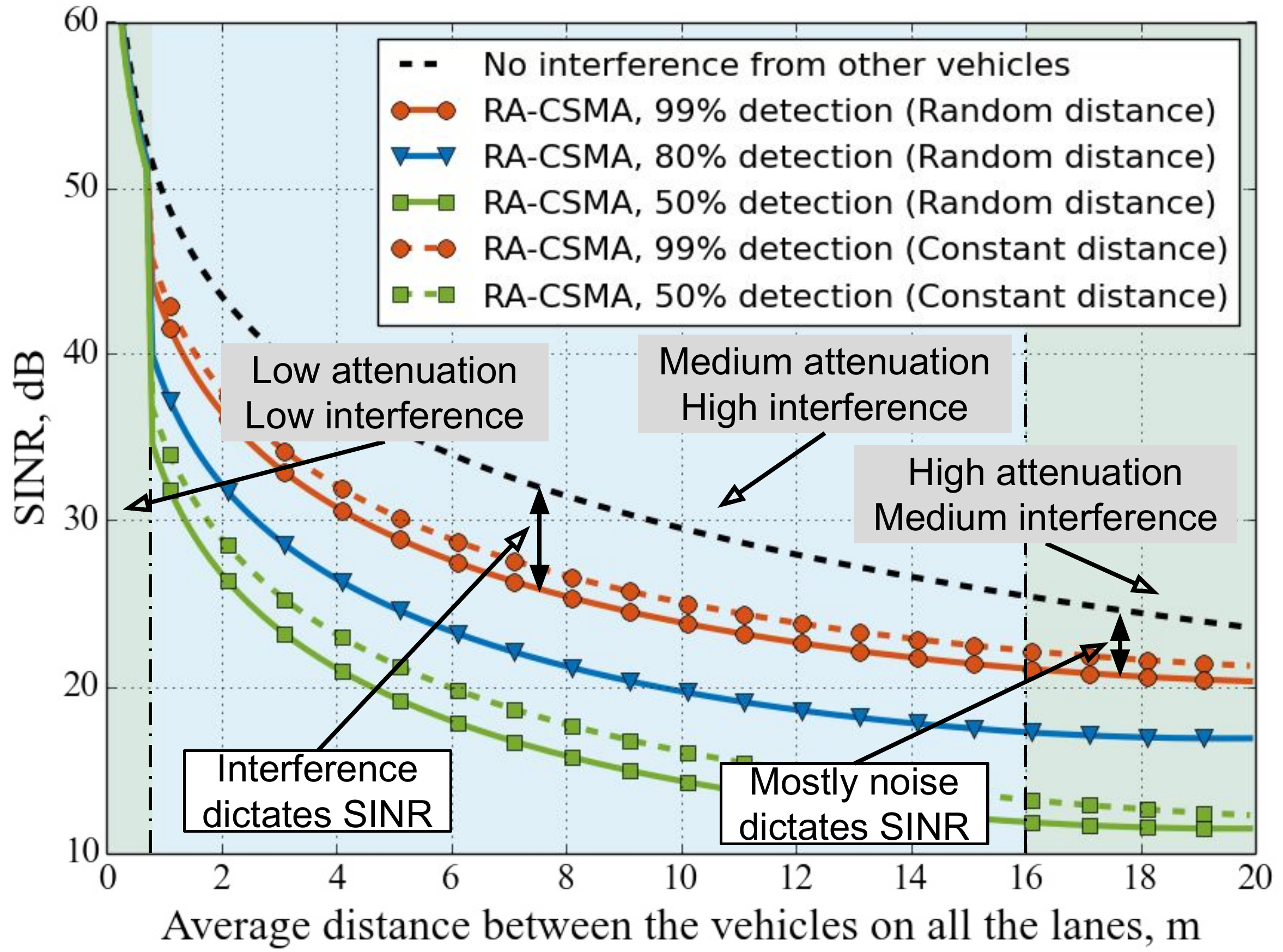}
\vspace{-3mm}
\caption{Impact of successful preamble detection on SINR performance.}
\label{fig:sinr}
\end{figure}

\textcolor{black}{Applying the results from Fig.~\ref{fig:sinr} to the radar sensing operation, we note that a connection between the estimated SINR values and the effective range of the radar sensing is complex, since the probability of successful obstacle detection at a particular distance is related not only to the SINR value but also to the specific obstacle shapes and material, the surrounding environment, and, importantly, the details of the radar implementation (type of the utilized radar, sampling frequency, filtering, and many more).} \textcolor{black}{Nevertheless, the improved SINR values -- as confirmed with Fig.~\ref{fig:sinr} -- contribute to the reliability of radar sensing, since higher SINR leads to better range and accuracy of the distance and velocity estimation~\cite{Gameiro2018}.}
 
Summarizing, the contributed channel access method offers decisive performance improvements in terms of \textcolor{black}{communications range, data rates, and SINR} in dense scenarios, thus contributing to better scalability of mmWave vehicular communications and radar sensing in massive deployments. In the following section, we discuss the current and prospective research directions related to V2X over mmWave and beyond.

\vspace{4mm}
\section{Towards Massive and Scalable V2X}

\subsection{Present and Future of V2X over mmWave and Beyond}

The contemporary V2X roadmap outlined by the 5G Public Private Partnership (5G-PPP) imposes gradually increasing requirements on the supporting communications technology in terms of its reliability, latency, and offered data rates. The envisioned 5G-PPP scenarios are grouped into phases that range from awareness driving, through cooperative driving, to, \textcolor{black}{ultimately}, synchronized driving, wherein vehicles are autonomously driven in a wide range of situations~\cite{my_white_paper_automotive_vision}.

Following the outlined use cases, novel approaches are required to achieve the desired tradeoff between the implementation complexity and the spectral efficiency. Specifically, for direct V2V communications, hybrid automatic repeat request (HARQ) operation over the 3GPP sidelink requires support of point-to-point communications at both PHY and MAC layers. In addition, the choice of the most beneficial locations and characteristics of the antenna arrays for hybrid V2X communications and radar sensing are currently under consideration.

\subsection{Harnessing Even Higher Frequencies}

Motivated by the continuously growing demands imposed by the amounts of data traffic to be transmitted over vehicular networks (up to $1$\,Tbit/hour per connected vehicle according to some predictions~\cite{heath_v2v_sensing}), the research community begins to contemplate the carrier frequencies beyond $100$\,GHz as a valuable asset for future V2X systems. Several radio communications stakeholders are already exploring the use of frequencies available in the high-mmWave/low-THz bands. The data rates of up to $10$\,Gbit/s over up to $850$\,m have been demonstrated at $120$\,GHz~\cite{research_plan_6}. An attractive potential of beyond-100\,GHz band has recently been recognized by IEEE: its IEEE 802.15.3d Task Group identified the $275$--$325$\,GHz band as the one feasible for extremely high access rates of up to $100$\,Gbit/s. Alongside with this trend, automotive radars are also moving higher up in frequency to improve their latencies and resolution of the obtained images. Recently, an automotive imaging radar solution at $150$\,GHz has been demonstrated~\cite{Jasteh2015}.

The path beyond $300$\,GHz requires the development of novel transceivers, which are capable of operating at these extreme frequencies. Such transceivers are required to be featured by adequately high power and sensitivity, while at the same time reach low noise figures -- all to overcome the inherently high pathloss at these frequencies. Additional band-specific difficulties emerge related to generation of high-power signals at (low-)THz frequencies, the so-called \emph{THz gap}, as well as increased attenuation of the radio waves due to \emph{molecular absorption}. In addition, the future high-mmWave/low-THz technology needs to manage the antenna systems with ultra-narrow beamwidth (few degrees at the maximum), thus challenging the current beamsteering and beamtracking implementations.

\textcolor{black}{Higher power consumption caused by immaturity of the discussed technology is another factor that limits commercial deployments of vehicular systems operating beyond $300$\,GHz. Researchers are now working on different technologies to tackle this problem, ranging from the state-of-the-art complementary metal-oxide-semiconductor (CMOS), through alternative semiconductor materials (e.g., III-V compound semiconductor), as well as novel materials, such as graphene. In addition, widely-employed orthogonal frequency division multiplexing (OFDM) features high peak-to-average power ratio (PAPR), which becomes an issue for terahertz-band channels. Hence, the community is exploring the applicability of novel waveforms to effectively harness this promising band.}

\subsection{Cellular-Assisted Network Management}

The further evolution of mmWave V2X requires their tighter integration into the emerging 5G architecture, where cellular systems may be employed to control the mmWave-based direct V2V and radar sensing more efficiently than any decentralized solution with only partial knowledge of the radio environment can do. The latter leads to the concept of cellular-assisted V2X (C-V2X). One of the main advantages of C-V2X is that it can address the V2X applications in an end-to-end manner within the same technology, which constitutes a scalable and future-proof solution. Also, as part of the 3GPP standards suite, C-V2X offers an evolution path from LTE to (beyond-)5G systems. Built-in mmWave radars can also exploit the benefits of cellular assistance as they may share the data between multiple cars and road infrastructure nodes to perform radio sensing in a distributed albeit more effective manner.

However, C-V2X functionality calls for new network management solutions to effectively orchestrate all of the available communications, data storage (caching), and computing capabilities. Particularly, the optimized balance levels of data pre-processing have to be determined between (i) exchanging all the sensed raw data among the interconnected vehicles, as one extreme, and a more conservative approach (ii) conducting the required sensing locally and only sharing the relevant post-processed knowledge, as another extreme. Design of these novel solutions mediates between vehicular, communications, and mobile computing communities, thus calling for their concerted effort to materialize the scalable, reliable, and high-rate vehicular communications of tomorrow.

\balance
\section{Conclusions}
\label{sec:conclusions}

The impending proliferation of interconnected autonomous vehicles will dramatically increase the intensity and depth of vehicular communications and radar sensing integration. The envisioned utilization of mmWave and even higher frequency bands will (i) provide abundant radio resources for high-resolution imaging, which is essential for accurate recognition of the road surface, cars, and other surrounding objects, as well as (ii) facilitate high-rate data exchange between connected and self-driving cars. At the same time, the adoption of wider spectrum and directional antenna systems alone will not automatically lead to a successful implementation of scalable vehicular systems, since the underlying interference and deafness issues challenge the reliability and performance of both communications and radar operations.

Along these lines, an emerging concept of unified vehicular communications and radar sensing contributed in this article becomes a promising candidate to make the two ``pairs of eyes'' of a vehicle \textcolor{black}{harmonized with} each other, thus paving the way to unprecedentedly massive deployments of interconnected smart cars. By going further and integrating other intelligent entities on and along the road (radio access infrastructures, networked road signs, connected pedestrians, video surveillance systems, etc.) via next-generation 3GPP NR, LTE, and IEEE technologies, the rapidly maturing automotive environment may ultimately evolve into a hyper-intelligent transportation system, by providing a more convenient, safe, and efficient driving experience.

\bibliographystyle{IEEEtran}
\bibliography{refs_3}

\section*{Authors' Biographies}

\textbf{Vitaly Petrov} (vitaly.petrov@tuni.fi) is a PhD candidate at the Laboratory of Electronics and Communications Engineering at Tampere University, Finland. He received the Specialist degree (2011) from SUAI University, St. Petersburg, Russia, as well as the M.Sc. degree (2014) from Tampere University of Technology. He is the recepient of Best Student Paper Award at IEEE VTC-Fall, Boston, USA, 2015 and Best Student Poster Award at IEEE WCNC, San Francisco, USA, 2017. Vitaly (co-)authored more than 30 published research works on terahertz band/mmWave communications, Internet-of-Things, nanonetworks, cryptology, and network security.

\textbf{Gabor Fodor} (gabor.fodor@ericsson.com) received the M.Sc. and Ph.D. degrees in electrical engineering from the Budapest University of Technology and Economics in 1988 and 1998 respectively. He is currently a master researcher at Ericsson Research and an adjunct professor at the KTH Royal Institute of Technology, Stockholm, Sweden. He was a co-recipient of the IEEE Communications Society Stephen O. Rice prize in 2018. He is serving as an Editor of the IEEE Transactions on Wireless Communications.

\textbf{Joonas Kokkoniemi} (joonas.kokkoniemi@oulu.fi) is a Postdoctoral Research Fellow with the Centre for Wireless Communications, University of Oulu. He received the B.Sc. (2011), M.Sc. (2012), and Dr.Sc. (2017) degrees from University of Oulu, Oulu, Finland. He was a Visiting Researcher with Tokyo University of Agriculture and Technology, Japan (2013) and a Visiting Researcher with State University of New York at Buffalo, USA (2017). Joonas's research interests include THz band and mmWave channel modeling and communication systems.

\textbf{Dmitri Moltchanov} (dmitri.moltchanov@tuni.fi) received the M.Sc. and Cand.Sc. degrees from the St. Petersburg State University of Telecommunications, Russia, in 2000 and 2003, respectively, and the Ph.D. degree from the Tampere University of Technology in 2006. Currently he is University Lecturer with the Laboratory of Electronics and Communications Engineering, Tampere University, Finland. He has (co-)authored over 150 publications. His current research interests include 5G/5G+ systems, ultra-reliable low-latency service, industrial IoT applications, mission-critical V2V/V2X systems and blockchain technologies.

\textbf{Janne Lehtom{\"a}ki} (janne.lehtomaki@oulu.fi) is an Adjunct Professor with the Centre for Wireless Communications, University of Oulu. He received the M.Sc. (1999) and the Ph.D. (2005) in telecommunications from University of Oulu. His research interests are in terahertz wireless communication, channel modeling, IoT, and spectrum sharing. Janne co-authored the winner of the Best Paper Award at IEEE WCNC 2012. He is an Editorial Board Member of Elsevier Physical Communication.

\textbf{Sergey Andreev} (sergey.andreev@tuni.fi) is an Assistant Professor in the Laboratory of Electronics and Communications Engineering at Tampere University, Finland. He received the Specialist degree (2006) and the Cand.Sc. degree (2009) both from St. Petersburg State University of Aerospace Instrumentation, St. Petersburg, Russia, as well as the Ph.D. degree (2012) from Tampere University of Technology. Sergey (co-)authored more than 100 published research works on wireless communications, energy efficiency, and heterogeneous networking.

\textbf{Yevgeni Koucheryavy} (evgeni.kucheryavy@tuni.fi) is a Full Professor in the Laboratory of Electronics and Communications Engineering of Tampere University, Finland. He received his Ph.D. degree (2004) from Tampere University of Technology. He is the author of numerous publications in the field of advanced wired and wireless networking and communications. He is Associate Technical Editor of IEEE Communications Magazine and Editor of IEEE Communications Surveys and Tutorials.

\textbf{Markku Juntti} (markku.juntti@oulu.fi) received a Dr.Sc. (Tech.) degree in electrical engineering from the University of Oulu, Finland, in 1997. He has been with the University of Oulu since 1992. In 1994--1995 he visited Rice University, Houston, Texas. He has been a professor of telecommunications at the University of Oulu since 2000. His research interests include communication and information theory, signal processing for wireless communication systems, and their application in wireless communication system design.

\textbf{Mikko Valkama} (mikko.valkama@tuni.fi) received his M.Sc. and D.Sc. degrees (both with honors) from Tampere University of Technology, Finland, in 2000 and 2001, respectively. In 2003, he worked as a visiting research fellow at San Diego State University, California. Currently, he is a Full Professor and Head of the Laboratory of Electronics and Communications Engineering at Tampere University. His research interests include radio communications, radio systems and signal processing, with specific emphasis on 5G and beyond mobile networks.

\end{document}